\documentclass[11pt]{article}
\usepackage{citesort}
\usepackage{amsmath}
\usepackage{amssymb}
\usepackage[dvips]{graphicx}
\usepackage{psfrag}
\usepackage{xspace}
\newlength{\picwidth}
\newlength{\figurewidth}
\newcommand{\kbar}{\makebox[0pt][l]{\hspace{0.2ex}\rule[1.25ex]{0.8ex}{0.3pt}}k}
\newcommand{\unit}[1]{\,\mathrm{#1}}
\newcommand{\ks}{k}
\newcommand{\etal}{\textit{et al}\xspace }
\begin{document}
\title{Experimental Evidence for the Role of Cantori as Barriers in a Quantum System}

\author{K. M. D. Vant, G. H. Ball, H. Ammann and N. L. Christensen,\\ Department of Physics, University of Auckland, Private Bag
92019,\\ Auckland, New Zealand}

\maketitle

\begin{abstract}
We investigate the effect of cantori on momentum diffusion in a
quantum system. Ultracold caesium atoms are subjected to a
specifically designed periodically pulsed standing wave. A
cantorus separates two chaotic regions of the classical phase
space. Diffusion through the cantorus is classically predicted.
Quantum diffusion is only significant when the classical
phase-space area escaping through the cantorus per period greatly
exceeds Planck's constant. Experimental data and a quantum
analysis confirm that the cantori act as barriers.
\end{abstract}
\section{Introduction}
\label{intro} The study of nonintegrable Hamiltonian systems has
contributed much to the understanding of dynamics in both the
classical and quantum domains. Classical phase space can contain
the rich structure of resonances, Kolmogo\-rov-Arnol'd-Moser (KAM)
tori, cantori and regions of stochasticity. Knowing quantum
mechanics to be the correct description of physical phenomena has
motivated physicists to search for the signatures of these
classical structures within the quantum regime. Hypersensitivity
to initial conditions, a hallmark of classical chaos, is
noticeably absent with quantum mechanics. The study [1-2] and
experimental observation of \emph{dynamical localization} [3-4]
dramatically displayed how quantum mechanics can eliminate
stochastic diffusion. There has always been the hope that a clear
transition between quantum and classical physics could be
achieved, but whereas one can always theoretically allow
$\hbar\rightarrow 0$ and realize classical behavior, $\hbar$'s
small but nonzero value always exists in the laboratory. With
$\hbar\neq 0$ there is a clear distinction between the properties
of classical chaos and its associated quantum counterpart.

The presence of KAM tori and cantori in the classical phase space
are predicted to influence the corresponding quantum system. If a
KAM boundary is unbroken it will prohibit classical diffusion
through it, but quantum mechanical tunneling across the barrier is
possible. When interaction terms in the perturbing Hamiltonian are
sufficiently large so as to break up the boundary and create a
cantorus or \emph{turnstile}, classical particles will quickly
diffuse through that cantorus but the quantum wavefunction will be
inhibited [5-8]. Given a periodic perturbing Hamiltonian, a
heuristic model proposes that when the classical phase-space area
escaping through the cantorus each period is $\sim\hbar$  then
quantum diffusion is constrained [7-8]. Even though the barrier
has been broken, the quantum wavefunction still appears to
\emph{tunnel} through the cantorus. The quantum system somehow
senses the cantorus, resulting in a diminished probability for
penetration of the barrier than that predicted classically. The
structure of this paper is as follows. In
Section~\ref{exptdetails} we introduce the experimental details of
our study of cantori constraining the diffusion of quantum
particles. The results of our experiment, along with our classical
and quantum calculations, are presented in Section~\ref{results}.
We examine the difference between dynamical localization and
cantori localization in Section~\ref{cantorilocalization}.
Finally, a summary is contained in Section~\ref{discussion}.
\section{Experimental Details}
\label{exptdetails}
A unique and pristine environment for studying quantum chaos is
achieved through the use of laser cooled neutral atoms within a
modulated standing wave of light [3-4,9-11]. The observation of
dynamical localization in the atomic optics realization of the
$\delta$-kicked rotor (DKR) [9-11] provided an important
experimental link to the most studied system in Hamiltonian chaos.
In this present paper we again subject our ultracold caesium atoms
to a pulsed standing wave of near resonant light. A periodic
pulsed potential of finite width (as used in the DKR experiments
[9-11]) produces a KAM boundary, which becomes more noticeable for
longer pulse widths. However, a train of single pulses is not the
best system for observing diffusion through cantori as the
classical phase space outside the boundary is not strongly chaotic
and contains many regular regions. We therefore subjected our
atoms to a train of double pulses. The pulse train consists of two
rectangular pulses of duration $\tau_{p}=1.25\unit{ms}$ with their
leading edges separated by $\tau_{s}=2.5\unit{ms}$. The double
pulses occur at every $T=25\unit{ms}$. Fig.\ 1(a) displays our
double pulse train. Adopting the notation used in [9-11], we can
write the dimensionless form of the Hamiltonian as,
\begin{equation}
H=\frac{\rho^{2}}{2}-k\cos\phi\sum_{l=1}^{N}f\left(t-l\right)
\end{equation}
where the $f(t)$ specifies the temporal shape of the pulses. For
an infinite train of kicks, the Hamiltonian can be written as,
\begin{equation}
H=\frac{\rho^{2}}{2}-k\sum_{r=-\infty}^{\infty}a_{r}\cos\left(\phi-2\pi rt\right)
\end{equation}
where $a_{0}=1/10$ and
$a_{r\neq0}=[\sin\left(3r\pi/20\right)-\sin\left(r\pi/20\right)]/(r\pi)$.
The classical interpretation of Eqn.\ (2) shows that the
fundamental resonances are located at $\rho=2\pi r$. The benefit
of using this type of pulsed system is illustrated in Fig.\ 1(b)
where the magnitudes of the $a_{r}$ terms are displayed. The KAM
boundary, located at $\rho=10\pi$ corresponds to the zero energy
of the $r=5$ term. Note that there is a relatively large amount of
energy for $r>5$. One can see in the Poincare sections displayed
in Fig.\ 2 that the $\rho=10\pi$ cantorus separates two strongly
chaotic regions; the boundary is clearly visible for $\ks=70$ but
is broken up into a chain of tiny islands beyond the resolution of
the simulation when $\ks=300$. Another KAM boundary at
$\rho=30\pi$ provides the upper boundary to the second region, and
remains unbroken for all kick strengths used in our experiment. We
prepare our atoms so that they initially lie within the
$\rho=10\pi$ cantori, and we monitor their subsequent evolution
through the barrier.

In order to establish the connection between the dimensionless
parameters used above and those that can be achieved in a
laboratory, we consider an atom (transition frequency
$\omega_{0}$) suspended in a standing wave of near resonant light
(frequency $\omega_{l}$). Under the assumption of a large detuning
compared to the Rabi frequency, the atoms' excited-state amplitude
can be adiabatically eliminated. The resulting Hamiltonian
governing the \emph{coherent} time evolution reads [9-11]
\begin{equation}
H=\frac{p^{2}}{2m}-\frac{\hbar\Omega_{eff}}{8}\cos\left(2k_{l}x\right)\sum_{q=1}^{N}f\left(t-qT\right)
\end{equation}
where
$\Omega_{eff}=\Omega^2\left(s_{45}/\delta_{45}+s_{44}/\delta_{44}+s_{43}/\delta_{43}\right)$
and $\Omega/2$  is the resonant Rabi frequency corresponding to a
single beam. The terms in brackets take account of the different
dipole transitions between the relevant hyperfine levels in
caesium $\left(F=4\rightarrow F'=5,4,3\right)$. The $\delta_{4j}$
are the corresponding detunings and, assuming equal populations of
the Zeeman sublevels, the numerical values for the $s_{4j}$ are
$s_{45}=11/27$, $s_{44}=7/36$, $s_{43}=7/108$. The function $f(t)$
represents the shape of our double kicks, discussed above. The
dimensionless Hamiltonian is then recovered with  $\phi=2k_{l}x$,
$\rho=2k_{l}Tp/m$, $t'=t/T$ and
$H'=\left(2k_{l}^{2}T^{2}/m\right)H$; the primes are subsequently
dropped. The kick strength is $\ks=\Omega_{eff}\omega{R}T^{2}$,
and $\omega_{R}=\hbar k_{l}^{2}/2m$ is the recoil frequency. The
quantum features of the DKR enter through the commutation relation
$[\phi,\rho]=i\kbar$, where $\kbar=8\omega_{R}T$. The relationship
between momenta is $n=p/\left(2\hbar k_{l} \right)=\rho/\kbar$.

Explicit details of the experimental set-up can be found in [11].
Approximately $10^{5}$ Cs atoms are initially trapped and cooled
in a magneto-optic trap (MOT) to a temperature of $10\unit{\mu
K}$. The position distribution of the trapped atoms has a FWHM of
$200\unit{\mu m}$. After trapping and cooling the magnetic field
gradient, the trapping beam and the repumping beam are turned off,
leaving the atoms in the $F=4$ ground state. The atoms are then
subjected to a modulated periodic potential from a third laser
diode. This beam passes through an acousto-optic modulator (AOM),
is collimated to a measured waist ($1/e^{2}$ intensity radius) of
$0.95\unit{mm}$, and is then retroreflected from a mirror outside
the vacuum cell to generate the one-dimensional potential. The
maximum Rabi frequency at the MOT center is
$\Omega/2\pi=100\unit{MHz}$, for a total power of $10\unit{mW}$.
The finite widths of the kicking beam waist and the atomic cloud
result in a narrow distribution of $\ks$ with rms spread of $6\%$
and $\ks_{mean}\approx0.94\ks_{max}$ ($\ks_{max}$ the value on the
beam axis). In the following, when specifying $\ks$, this always
refers to $\ks_{mean}$. The kick strength $\ks$ was varied by
adjusting $\Omega$ while maintaining a detuning of
$\delta_{45}/2\pi=2.8\unit{GHz}$ to the blue. For our detuning the
spread in coupling strengths created by differing AC Stark shifts
of the different magnetic sublevels was below $1\%$. To measure
the atomic momentum distribution we use a time-of-flight technique
with a `freezing molasses' [9-11] with an expansion time of
$12\unit{ms}$.
\section{Results from calcualtions and the experiment}
\label{results}
The purpose of the experiment discussed in this paper is to
examine the diffusion of particles through the barriers at
$\rho=\pm10\pi$. Initially our Cs atoms are prepared in the MOT
with a temperature of $10\unit{\mu K}$, or a variance of
$\sigma_{\rho}=9.2$ in terms of the dimensionless momentum $\rho$.
This places the initial momentum distribution within the cantori.
After a number of kick cycles of sufficient strength the resulting
atomic momentum distribution quickly resembles that displayed in
Fig.\ 3. The effects of the cantori are clearly discernible. For
this example we had $\ks=310$ and $N=56$. Note the wings in the
distribution outside of the $\rho=\pm10\pi$ cantori, but before
the $\rho=30\pi$ boundary. For comparison we display the predicted
distributions for quantum and classical calculations. The
asymmetry in the measured lineshape around $\rho=0$ is due to
fringes in the freezing molasses. The lineshapes are consistent in
appearance over two or three days, at which point the entire
optical system is typically re-optimized. Fig.\ 4 displays the
temporal evolution of the lineshapes for $\ks=300$. The traces
correspond to $N=3,6,44$ and $68$. There is an initial spreading
of the lineshape, followed by a subsequent saturation with
shoulders appearing on the lineshape at the location of the
$\rho=\pm10\pi$ cantori.

In keeping with the work of Geisel \etal [5-6] we measured and
calculated the percentage of particles that would cross the
$\rho=\pm10\pi$ cantori as a function of kick number and kick
strength. All our measurements were at a fixed ``Planck's
constant'' of $\kbar=2.6$; the consequences of a variation of
$\kbar$ in this double pulse system will be the subject of future
work. In this present work we also purposely avoid the influence
of spontaneous emission. The study of how decoherence introduced
via spontaneous emission can affect the ability of particles to
flow through the cantori is intriguing, and will be presented in a
forthcoming publication. In this present study we accounted for
spontaneous emission by including it in our quantum computation,
and comparing it to a computation where the effect is absent. We
can therefore confirm that our experimental results are from a
regime where the influence of spontaneous emission is minimal, but
still noticeable. Our quantum analysis is based on a density
matrix calculation. Spontaneous emission is included via the
inclusion of an interaction term, $H_{int}=-\zeta
u\kbar\phi\sum_{l=1}^{N}\delta\left(t-l\right)$  to the
Hamiltonian (1), where $\zeta$ is either $0$ or $1$,
$\langle\zeta\rangle=\eta=$ probability for spontaneous emission
per double kick, and $u\kbar$ is the recoil momentum projected
onto the kicking beam axis ($u$ chosen randomly on the interval
$[-1,1]$).

The influence of KAM boundaries on the momentum distribution of
atoms in similar experiments has been previously observed [4]. The
Austin group has also recently carried out a systematic
investigation of the finite-time pulsed rotor system [12], with
single kicks. By increasing the duration of the pulses they bring
in the KAM boundary and observe a reduction in the energy of the
atomic distribution.

There exists a heuristic explanation for the inhibition of quantum
diffusion through partial barriers and turnstiles [7-8]. When the
flux per period of phase space area escaping through classical
cantorus is comparable to or less than Planck's constant, then
quantum diffusion is suppressed and penetration through the
barrier is only via tunneling. Our experimental observations and
theoretical analysis support this hypothesis. In Fig.\ 5 we
display the results of quantum and classical analyses that predict
the percentage of atoms that will be found outside the
$\rho=\pm10\pi$ cantori as a function of kick number for various
kick strengths; both the classical and quantum calculations omit
spontaneous emission. The initial distribution corresponds to
$10\unit{\mu K}$ or a variance of $\sigma_{\rho}=9.2$, or $99.9\%$
of the atoms within the $\rho=\pm10\pi$ cantori. Our computer
simulations reveal that the KAM boundary is broken at $\ks\leq50$.
With the barrier broken the classical particles will eventually
spread themselves uniformly between the $\rho=\pm30\pi$ barriers,
thereby giving a probability to be found outside the
$\rho=\pm10\pi$ cantori of $2/3$. For $\ks=310$ the classical
particles reach this equilibrium in about $70$ kicks.

We have numerically calculated the classical phase space flux
through the cantorus per kicking cycle. We assume the particles
are uniformly spatially distributed, but with initial momentum of
$\rho=10\pi$. Referring to the kick strengths displayed in Fig.\
5, we find that the phase space area per period moving across the
cantorus to larger momenta is proportional to $\ks^{2}$, with
values of $5.4\kbar$, $3.4\kbar$ and $0.88\kbar$ for $\ks = 310,
240$, and $120$ respectively. The ability of the cantori to
constrain the quantum particles is clearly displayed. The quantum
system quickly reaches its own equilibrium, with significantly
fewer atoms outside the cantori. Only for kick strengths around
$\ks=1200$ does the quantum system begin to mimic the classical,
with the probability for penetration of the cantori exceeding
$60\%$ and the phase space area per period moving across the
cantorus at $31\kbar$.

The experimental results for our diffusion experiment are shown in
\mbox{Fig.\ 6}. Displayed are the measured probabilities of
finding our Cs atoms outside the $\rho=\pm10\pi$ cantori. Also
shown are the results of quantum analyses that include the
spontaneous emission rates for the experimental parameters. The
spontaneous emission rate per kicking cycle is $\eta = 0.021,
0.017$ and $0.008$ for $\ks = 310, 240$ and $120$ respectively. We
see good agreement between the measured and calculated
probabilities. For strong kick strengths the final measured
probabilities are noticeably smaller than those predicted
classically; for $N\approx70$ we measure $37\%$ and $12\%$ for
$\ks = 310$ and $240$ respectively, whereas the classical
prediction gives $62\%$ and $48\%$. The broken cantori still
dramatically restrict the movement of atoms through the
classically broken barrier. By blocking our retroreflected beam we
experimentally confirm that our small spontaneous emission rates
create minimal heating and do not produce any movement of the
atoms through the cantori.

The resolution of our CCD ($19$ pixels/mm), coupled with our
$12\unit{ms}$ expansion time, allowed us to determine the position
of the momentum line to an accuracy of $\Delta\rho=\pm0.8$. At our strongest kick
strength of $\ks=310$ this results in an uncertainty in our measured
probability of $\pm4\%$. We repeat our experiment a number of times
during each experimental session, and the resulting spread in our
measured values is $\pm4\%$.
\section{Cantori localization and dynamical localization}
\label{cantorilocalization}
Different process can contribute to a quantum system's inhibition
to diffusion. Dynamical localization is certainly the reason for
the elimination of momentum diffusion in the kicked rotor [1-2]
and its atomic optics realization [9-11] at sufficiently high kick
strengths; we note that others [13] disagree with this conclusion
for the experiment described in Refs.\ [3-4]. Recent theoretical
work on discontinuous quantum systems displays dynamical
localization and cantori inhibition of diffusion in the same
system [14]. Previously other studies also focused on the
transition from cantori localization to dynamical localization
[15]. For the experiment described in this present paper the
cantori also act as barriers. Whereas dynamical localization
results in an exponential lineshape for the momentum distribution
in the $\delta$-kicked rotor, the quantum version of our double
kick system quickly settles into the type of distribution that is
displayed in Figs.\ 3 and 4. Our measurements and quantum analysis
show that the system's probability for penetration of the barrier
saturates at values far below the classical prediction.

The effect of dynamical localization is universal, and appears in
a broad class of systems. To some extent it will also contribute
to the restriction of momentum diffusion in our double kick
system. In order to more closely examine dynamical and cantori
localization we compare the double kicked system to one with a
single kick, but with the same total pulse area. In the following,
the kick strength is $k=300$, while the single kick has a pulse
length of $\tau=2.5\unit{\mu s}$, and pulse period $T=25\unit{\mu
s}$. For the single kick system there will be no $\rho=\pm10\pi$
cantori, but KAM boundaries at $\rho=\pm20\pi$. \mbox{Fig.\ 7}
displays the results of a quantum calculation for the momentum
lineshapes after $50$ kick cycles. The calculation includes the
slight effect of spontaneous emission rate of $\eta=0.02$ per kick
cycle for a detuning of $2.8\unit{GHz}$. The single kick lineshape
is not perfectly exponential in shape, and is definitely
constrained by its $\rho=\pm20\pi$ KAM boundary. Similar
lineshapes for the finite pulse kicked rotor have been previously
experimentally studied [12]. One can clearly observe the different
shape for the double kick momentum distribution; the shoulders
develop at the location of the $\rho=\pm10\pi$ cantori. In Fig.\ 8
we have the experimentally observed momentum lineshapes for the
single and double kicked atoms. Once again the shoulders are
observed for the double kicked system at the $\rho=\pm10\pi$
cantori. The single kick momentum distribution is broad enough
such that its localization length would extend to the location of
the cantori at $\rho=\pm10\pi$. Hence the effect of these cantori
become important and noticeable in the double kick system.

If we model our kicked system as a kicked rotor then a prediction
can be made for the localization length one would expect from
dynamical localization. A kick strength of $\ks=300$, a single
kick pulse length of $\tau=2.5\unit{\mu s}$, and pulse period
$T=25\unit{\mu s}$ provides a classical stochasticity parameter
for the kicked rotor of $\kappa=30$ [10-11]. Dynamical
localization would predict a characteristic exponential
distribution $\sim\exp\left(-2|\rho|/l_{\rho}\right)$, with
localization length $l_{\rho}\approx\kappa^{2}/2\kbar\approx170$
[1]. This greatly exceeds the cantori at $\rho=\pm10\pi$ for the
double kicked system, hence the shoulders that develop are
evidence for cantori localization. For the single kicked system
one can observe a box-like distribution due to the $\rho=\pm20\pi$
boundaries, which are also smaller than the localization length.

For our parameters ($\ks=300$, $N=50$, $\delta=2.8\unit{GHz}$) the
single kick system saturates to a momentum distribution with an
additional $10\%$ of the total number of atoms in excess of
$\rho=\pm10\pi$. Fig.\ 9 compares the percentage of atoms that
cross the momentum value of $\rho=\pm10\pi$ for the single and
double kicks. This calculation includes the spontaneous emission
rate of $\eta=0.02$ per kick cycle. Also, the single kick system
has a `quantum break time' of approximately $N=7$, whereas the
double kicked system needs about $15$ kicks to come to
equilibrium. The $\rho=\pm10\pi$ cantori slow the initial
diffusion of atoms. The quick equilibrium for similar finite
length single pulses has been experimentally observed [12], while
the slow initial diffusion is clearly seen in our Fig.\ 6.
\section{Discussion}
\label{discussion}
Decoherence in the atomic optics manifestation of the kicked rotor
has recently been observed [10-11,16]. The decohering effect of
spontaneous emission is small but still noticeable in our present
data. Whereas a fully quantum system saturates and the percentage
of atoms found beyond the $\rho=\pm10\pi$ cantori remains
constant, spontaneous emission creates a slow drift in the system
toward the classical equilibrium configuration where $2/3$ of the
particles are outside these cantori. The coupling of a quantum
system to extraneous degrees of freedom, or the \emph{environment},
destroys quantum coherences. In these quantum chaos experiments
utilizing atomic optics the quantum dynamics become susceptible to
the decohering effects of spontaneous emission. The environment in
this case is the vacuum fluctuations. A heuristic model for
decoherence observed in our double kick system, along with further
experimental data, will be presented in a forthcoming publication.
The model is similar to that presented in [10-11]. However, from
an experimental view of our present data one can clearly observe
the slow drift of our quantum system towards the classical
equilibrium configuration.

By using a periodic train of double (finite width) pulses we
create a system with two strongly chaotic regions separated by a
cantorus. Using laser cooled atoms and a pulsed laser beam we have
a pristine environment for experimentally observing a quantum
system's inhibition to diffusion through cantori. Previous
theoretical work [5-8] supports the hypothesis that when the flux
of classical phase space through turnstiles is less than Planck's
constant then the movement of particles across the barrier is
restricted. Our measurements and calculations of the movement of
atoms through the cantorus support the conclusion that when the
classical phase space flux through the cantori per kicking cycle
is $\lesssim\kbar$, the system fails to see the holes in the
cantori [14]. The restriction of the quantum particle's momentum
is remarkable when one considers the underlying classical phase
space. For the kick strengths used in our experiment the
$\rho=\pm10\pi$ cantori are not visible in the Poincare section,
just an apparent region of stochasticity constrained by the
barriers at $\rho=\pm30\pi$. However, the effect of the cantori
remains significant for the quantum system.
\section*{Acknowledgements} This work was supported by the Royal Society of New
Zealand Marsden Fund and the University of Auckland Research
Committee.

\section*{References}
[1] D.L. Shepelyansky, Physica D 28, 103 (1987)\\{}
[2] S. Fishman, D.R. Grampel, and R.E. Prange, Phys. Rev. Lett.
49, 509 (1982)\\{}
[3] F.L. Moore, J.C. Robinson, C. Bharucha, P.E. Williams, and
M.G. Raizen, Phys. Rev. Lett. 73, 2974 (1994)\\{}
[4] J.C. Robinson, C. Bharucha, F.L. Moore, R. Jahnke, G.A.
Georgakis, Q. Niu, M.G. Raizen, and B. Sundaram, Phys. Rev. Lett.
74, 3963 (1995)\\{}
[5] T. Geisel, G. Radons, and J. Rubner, Phys. Rev. Lett. 57, 2883
(1986)\\{}
[6] T. Geisel and G. Radons, Physica Scripta 40, 340 (1989)\\{}
[7] R.C. Brown and R.E. Wyatt, Phys. Rev. Lett. 57, 1 (1986)\\{}
[8] R.S. MacKay and J.D. Meiss, Phys. Rev A  37, 4702 (1988)\\{}
[9] F.L. Moore, J.C. Robinson, C.F. Bharucha, B. Sundaram, and
M.G. Raizen, Phys. Rev. Lett. 75, 4598 (1995)\\{}
[10] H. Ammann, R. Gray, I. Shvarchuck and N. Christensen, Phys.
Rev. Lett. 80, 4111 (1998)\\{}
[11] H. Ammann, R. Gray, N. Christensen, and I. Shvarchuck, J.
Phys. B: At. Mol. Opt. Phys. 31, 2449 (1998)\\{}
[12] B.G. Klappauf, W.H. Oskay, D.A. Steck, and M.G. Raizen,
Physica D, in press (1998)\\{}
[13] M. Latka and B.J. West, Phys. Rev. Lett. 75, 4202 (1995), and
resulting comments, M.G. Raizen, B. Sundaram, and Q. Niu, Phys.
Rev. Lett. 78, 1194 (1997), S. Meneghini, P.J. Bardroff, E. Mayr,
and W.P. Schleich, Phys. Rev. Lett. 78, 1195 (1997), M. Latka and
B.J. West, Phys. Rev. Lett. 78, 1196 (1997)\\{}
[14] F. Borgonovi, Phys. Rev. Lett. 80, 4653 (1998)\\{} [15] G.
Radons, T. Geisel, and J. Rubner, Adv. Chem. Phys. 73, 891
(1989)\\{} [16] B.G. Klappauf, W.H. Oskay, D.A. Steck, and M.G.
Raizen, Phys. Rev. Lett. 81, 1203 (1998)\\{}
\begin{figure}[p]
\psfrag{aT}{\large{$\tau_{p}$}}\psfrag{dT}{\large{$\tau_{s}$}}
\psfrag{T}{\large$T$}
\psfrag{x}[ct][cB]{\Large $|\rho|/\pi$}
\psfrag{y}[cB][ct]{\raisebox{3mm}{\Large Relative magnitude}}
\psfrag{a}[c][c]{0} \psfrag{b}[c][c]{5} \psfrag{c}[c][c]{10}
\psfrag{d}[c][c]{15} \psfrag{e}[c][c]{20} \psfrag{f}[c][c]{25}
\psfrag{g}[c][c]{30} \psfrag{h}[c][c]{35} \psfrag{i}[c][c]{40}
\psfrag{j}[rB][rB]{0} \psfrag{k}[rB][rB]{0.1}
\psfrag{l}[rB][rB]{0.2} \psfrag{m}[rB][rB]{0.3}
\psfrag{n}[rB][rB]{0.4} \psfrag{o}[rB][rB]{0.5}
\psfrag{p}[rB][rB]{0.6} \psfrag{q}[rB][rB]{0.7}
\psfrag{r}[rB][rB]{0.8} \psfrag{s}[rB][rB]{0.9}
\psfrag{t}[rB][rB]{1.0}
\centering\scalebox{0.6}{\begin{minipage}{210mm}\begin{center}
\begin{minipage}{0.61\textwidth}\begin{center}(a)\\
\includegraphics[clip,width=0.99\textwidth,bb=78 107 456 287]{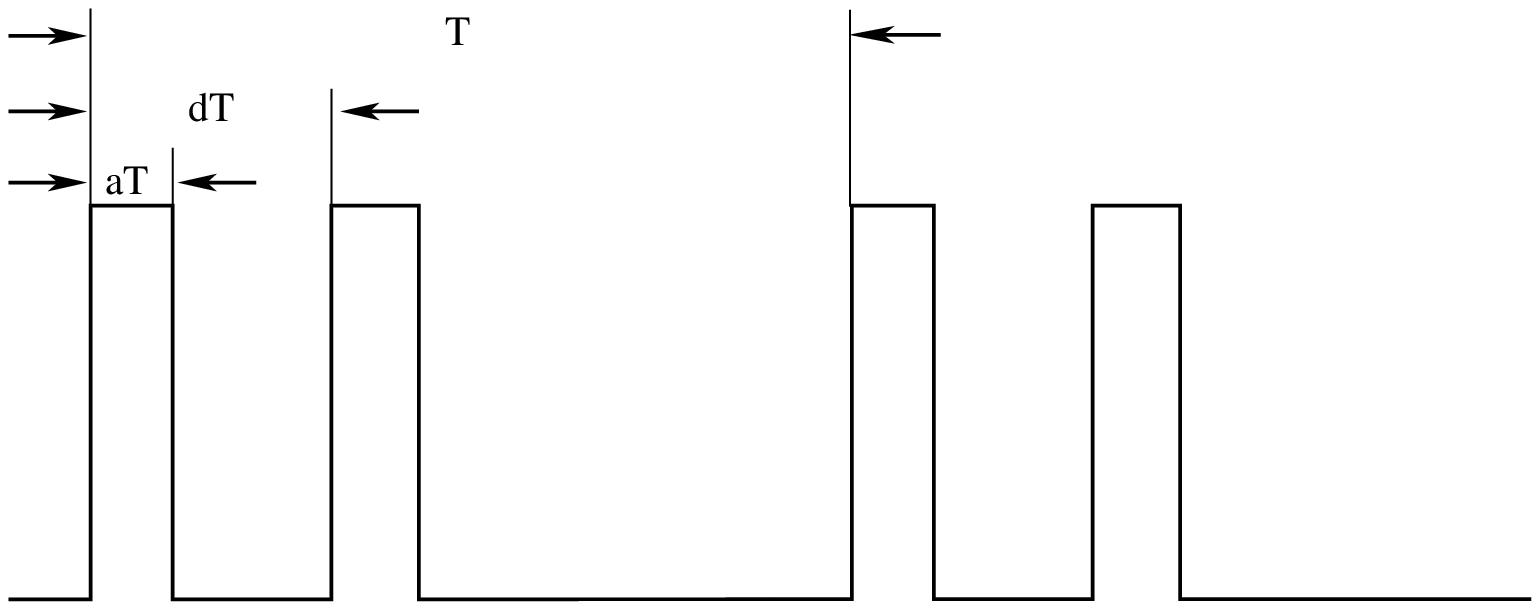}
\end{center}\end{minipage}\\[20pt]
\begin{minipage}{0.61\textwidth}\begin{center}(b)\\
\includegraphics[width=0.99\textwidth]{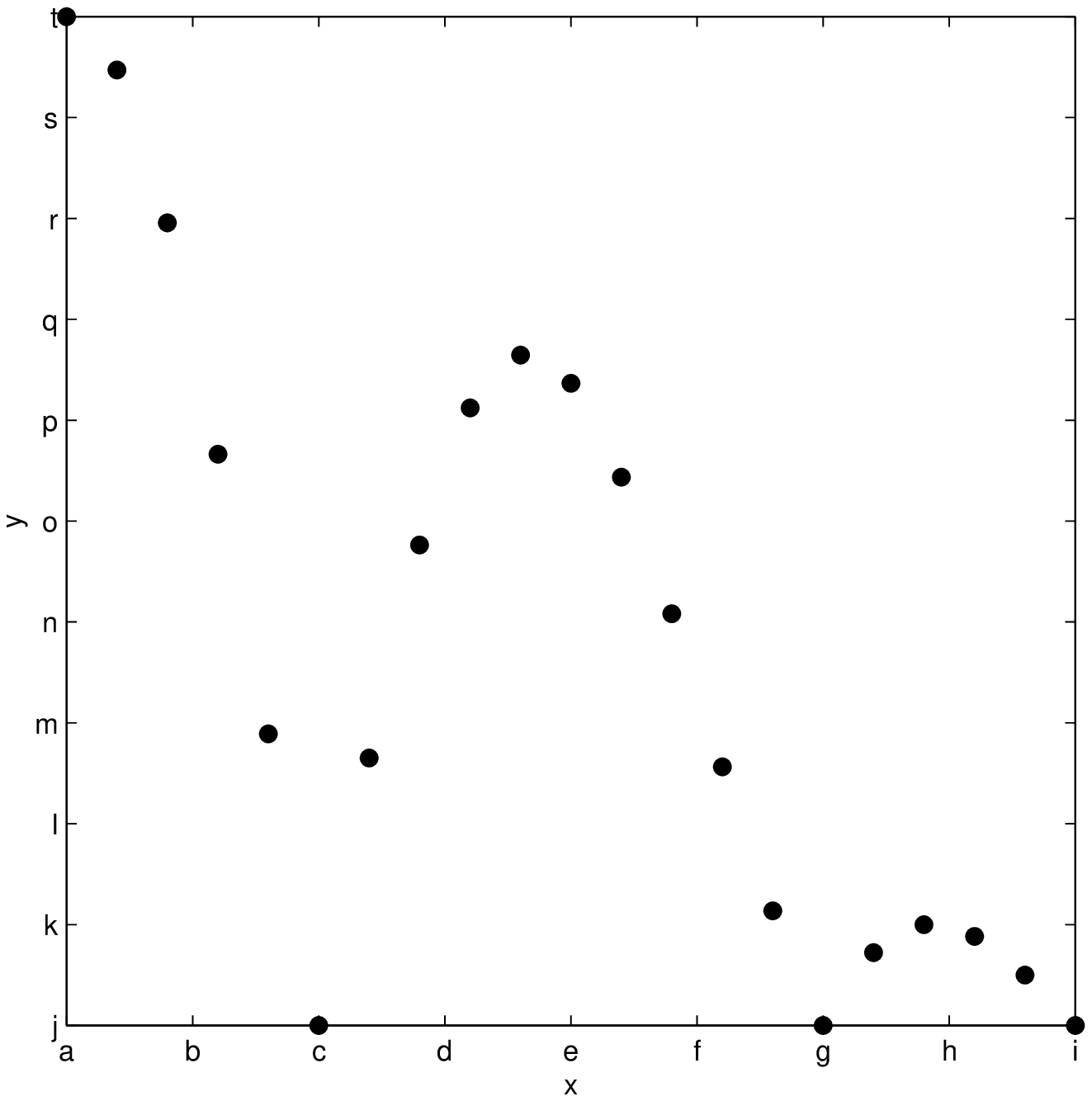}
\end{center}
\end{minipage}\end{center}\end{minipage}}
\caption{Time sequence (a) of the pulse, and the magnitude of the
Fourier components $a_{r}$ (b) used in the Hamiltonian, Eqn.\
(2).}
\end{figure}
\begin{figure}[p]
\begin{center}
\includegraphics[bb=140 120 360 600]{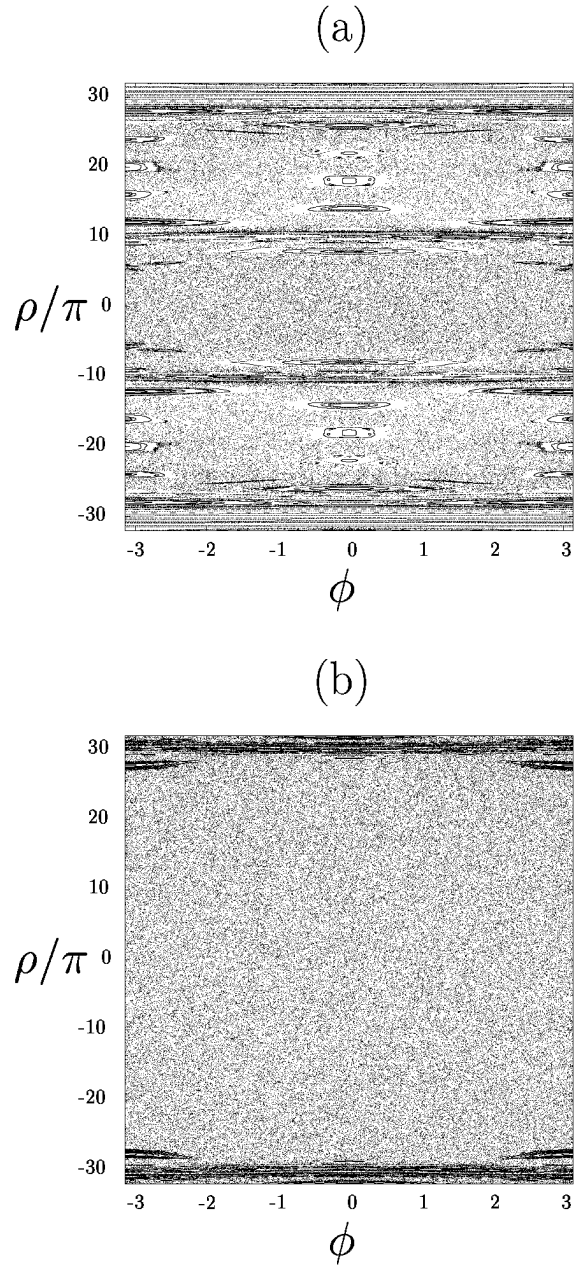}
\end{center}
\caption{Classical phase space for the relatively small kick
strength $\ks=70$ (a), with boundaries clearly visible at
$\rho=\pm10\pi$ and $\rho=\pm30\pi$. For the kick strength of
$\ks=300$ (b) the $\rho=\pm10\pi$ cantori have completely broken
up.}
\end{figure}
\begin{figure}[p]
\centering
\includegraphics[width=\figurewidth]{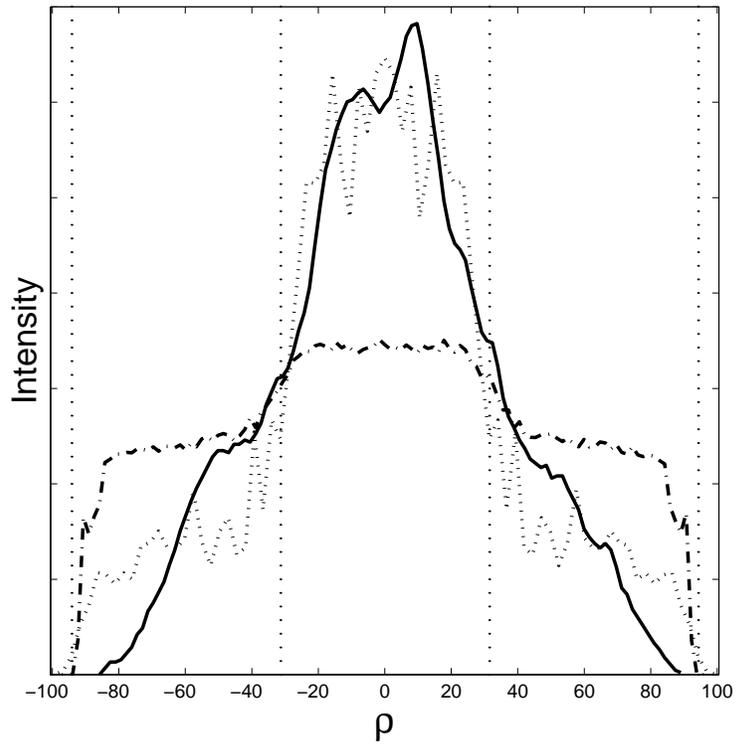}
\caption{An example of a measured momentum distribution (solid)
for kick strength $\ks=310$ and number of kicks $N=56$ , along
with that predicted via a quantum (dotted) or classical
(dot-dashed) analysis. The vertical lines correspond to the
location of the barriers at $\rho=\pm10\pi$ and $\rho=\pm30\pi$.}
\end{figure}
\begin{figure}[p]
\centering
\includegraphics[width=\figurewidth]{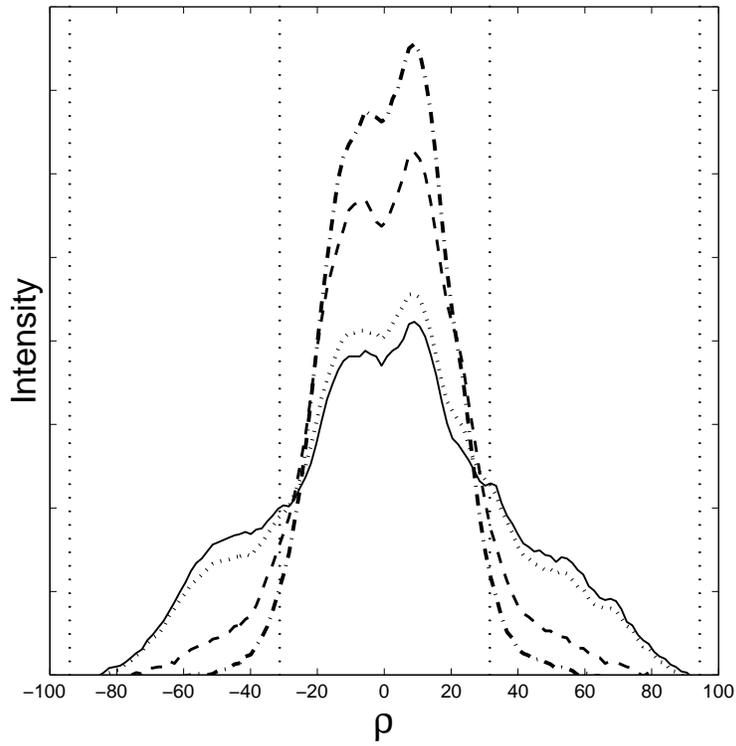}
\caption{Measured momentum distribution for kick strength
$\ks = 300$ as the number of kicks increases; $N = 3$ (dot-dashed
line), $N = 6$ (dashed line), $N = 44$ (dotted line), and $N = 66$
(solid line). The vertical lines correspond to the location of the
barriers at $\rho=\pm10\pi$ and $\rho=\pm30\pi$.}
\end{figure}

\begin{figure}
\psfrag{a}[c][c]{0} \psfrag{b}[c][c]{10} \psfrag{c}[c][c]{20}
\psfrag{d}[c][c]{30} \psfrag{e}[c][c]{40} \psfrag{f}[c][c]{50}
\psfrag{g}[c][c]{60} \psfrag{h}[c][c]{70} \psfrag{k}[r][r]{0}
\psfrag{n}[r][r]{10} \psfrag{o}[r][r]{20} \psfrag{p}[r][r]{30}
\psfrag{q}[r][r]{40} \psfrag{r}[r][r]{50} \psfrag{s}[r][r]{60}
\psfrag{t}[r][r]{70}
\psfrag{x}[ct][cB]{\large Number of double kicks}
\psfrag{y}[cB][ct]{\raisebox{4mm}{\large \% of particles outside
$\rho=\pm10\pi$ cantori}}
\centering
\includegraphics[width=\figurewidth]{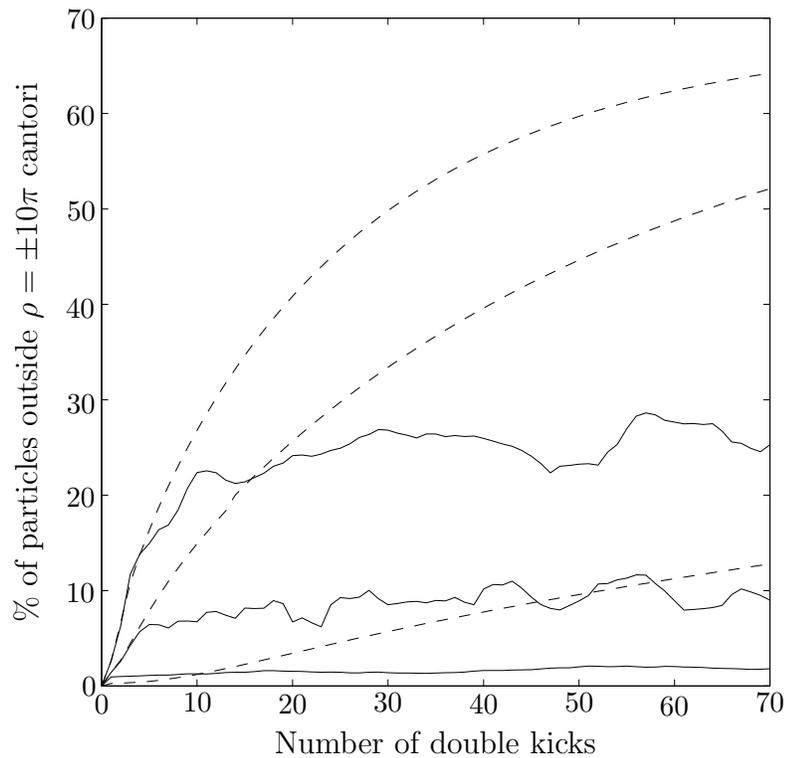}
\caption{The probability for finding the particles outside
the $\rho=\pm10\pi$ cantori, versus kick number as calculated via
classical (dashed) and quantum (solid) analyses for kick strengths
of $\ks = 310$ (upper), $240$ (middle), and $120$ (lower).}
\end{figure}

\begin{figure}[p]
\psfrag{a}[c][c]{0} \psfrag{b}[c][c]{10} \psfrag{c}[c][c]{20}
\psfrag{d}[c][c]{30} \psfrag{e}[c][c]{40} \psfrag{f}[c][c]{50}
\psfrag{g}[c][c]{60} \psfrag{h}[c][c]{70} \psfrag{k}[r][r]{0}
\psfrag{n}[r][r]{10} \psfrag{o}[r][r]{20} \psfrag{p}[r][r]{30}
\psfrag{q}[r][r]{40} \psfrag{r}[r][r]{50} \psfrag{s}[r][r]{60}
\psfrag{t}[r][r]{70}
\psfrag{x}[ct][cB]{\large Number of double kicks}
\psfrag{y}[cB][ct]{\raisebox{4mm}{\large \% of particles outside
$\rho=\pm10\pi$ cantori}}
\centering
\includegraphics[width=\figurewidth]{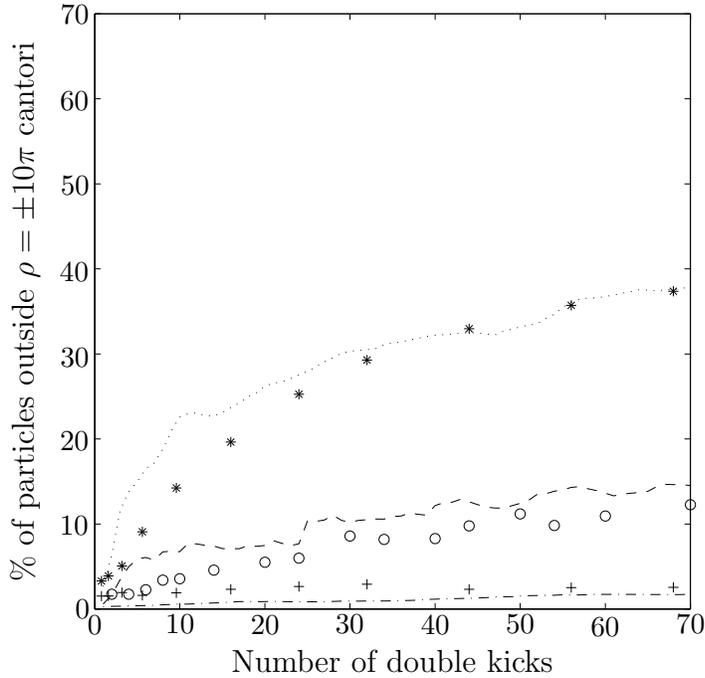}
\caption{The probability for finding the particles outside the
$\rho=\pm10\pi$ cantori, versus kick number for experimental (data
points) and a quantum simulation (lines) that includes spontaneous
emission for kick strengths of $\ks = 310$ (* and dotted line),
$240$ (  and dashed line) and $120$ (+ and dot-dashed line). The
position of the $\rho=\pm10\pi$ momentum line is determined to an
accuracy of $\Delta\rho=\pm0.8$, resulting in an uncertainty in
our measured probability of $\pm4\%$. Multiple repetition of our
experiment provides a spread in the measured values of $\pm4\%$.}
\end{figure}
\begin{figure}
\centering
\includegraphics[width=\figurewidth]{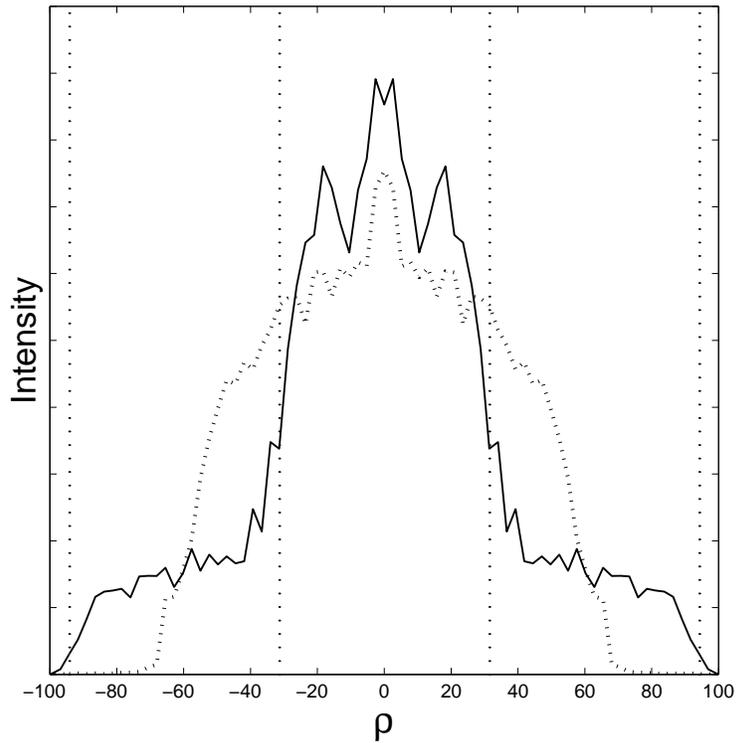}
\caption{An example of the calculated momentum
distribution for kick strength $\ks = 300$ and number of kicks
$N=50$ for the double (solid line) and single (dotted line) kicked
systems. The vertical lines correspond to the location of the
barriers at $\rho=\pm10\pi$ and $\rho=\pm30\pi$. The calculation
includes a spontaneous emission rate of $\eta=0.02$ per pulse
period.}
\end{figure}
\begin{figure}
\centering
\includegraphics[width=\figurewidth]{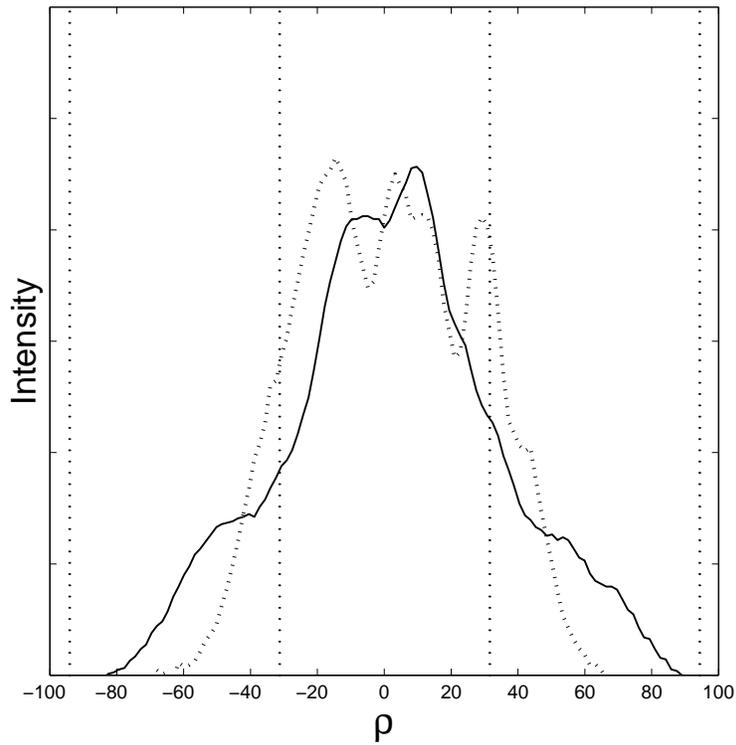}
\caption{An example of a measured momentum distribution
for kick strength $\ks = 300$ and number of kicks $N=50$  for the
double (solid line) and single (dotted line) kicked systems at a
detuning of $2.8\unit{GHz}$. The vertical lines correspond to the
location of the barriers at $\rho=\pm10\pi$ and $\rho=\pm30\pi$.}
\end{figure}
\begin{figure}
\psfrag{x}[ct][cB]{\Large Number of kicks/double kicks}
\psfrag{y}[cB][ct]{\Large \% outside $\rho=10\pi$}
\centering
\includegraphics[width=\figurewidth]{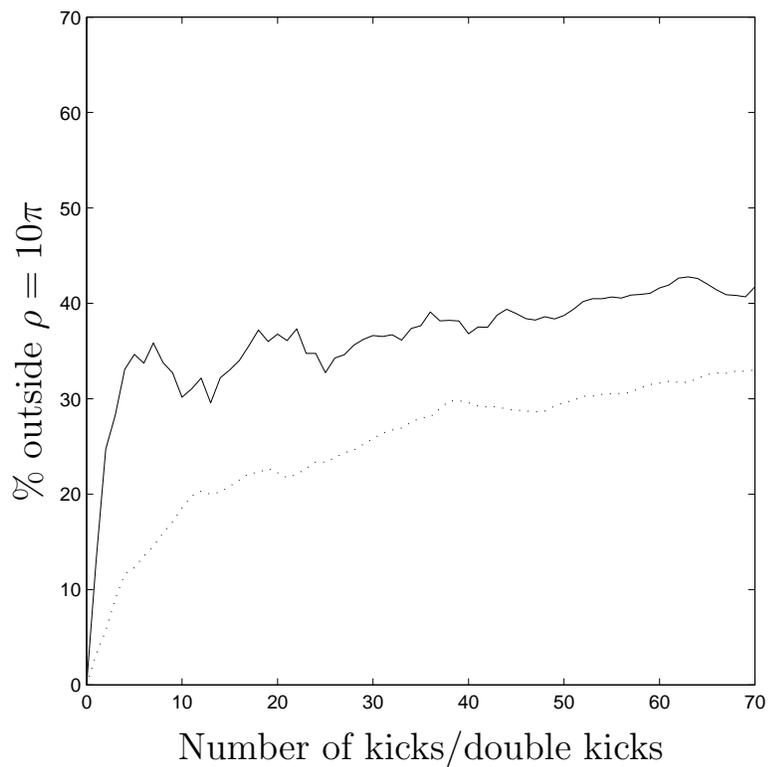}
\caption{The calculated probability for finding particles in
excess of $\rho=\pm10\pi$, versus kick number, as calculated for a
kick strength of $\ks = 300$ and spontaneous emission rate of
$\eta=0.02$ per pulse period. The solid line is the single kick,
while the dotted line is the double kick.}
\end{figure}
\end{document}